# Spontaneous and Stimulated Raman Scattering near Metal Nanostructures in the Ultrafast, High-Intensity regime

M. Scalora[1], M. A. Vincenti[1,2], D. de Ceglia[1,2], M. Grande[3], J. W. Haus[2,4]

[1] *Charles M. Bowden Research Center, AMRDEC, RDECOM, Redstone Arsenal, Alabama 35898-5000, USA*

[2] *National Research Council - AMRDEC, Charles M. Bowden Research Center, Redstone Arsenal - AL, 35898 – USA*

[3] *Dipartimento di Ingegneria Elettrica e dell'Informazione (DIEI), Politecnico di Bari, Via Re David 200, 70126 Bari – Italy*

[4] *Electro-Optics Program, University of Dayton, 300 College Park, Dayton, OH 45469*

## ABSTRACT

The inclusion of atomic inversion in Raman scattering can significantly alter field dynamics in plasmonic settings. Our calculations show that large local fields and femtosecond pulses combine to yield: (i) population inversion within hot spots; (ii) gain saturation; and (iii) conversion efficiencies characterized by a switch-like transition to the stimulated regime that spans twelve orders of magnitude. While in Raman scattering atomic inversion is usually neglected, we demonstrate that in some circumstances full accounting of the dynamics of the Bloch vector is required.

In a recent paper we outlined a fresh theoretical approach to describe Raman scattering near metal nanostructures based on a time domain, fast Fourier transform beam propagation method [1]. One assumes that the portion of the medium that makes up the generic nanostructure (metal, dielectric, and/or metal-dielectric) is described classically by a combination of Drude-Lorentz oscillators to account for free and bound electron contributions that generate a macroscopic polarization. The surrounding Raman-active medium is described quantum mechanically by a set of Bloch equations that give rise to an additional contribution to the macroscopic polarization. The material equations of motion are then combined and solved simultaneously with Maxwell's equations in the time domain. The approach departs from usual methods used to describe processes like surface enhanced Raman scattering (SERS) in that it deploys a full set of nonlinear, coupled equations to solve for the dynamics of pump, Stokes, and anti-Stokes fields, and it also accounts for pump depletion. The use of equations of motion for all fields and the calculation of emitted energies supplant estimates of Raman gain that come about by a mere utilization of the linear field modes to predict a Raman enhancement factor [2, 3], usually defined as $G_S = \frac{<|E^{local}_{\omega_p}|^2><|E^{local}_{\omega_s}|^2>}{<|E^{inc}_{\omega_p}|^4>}$ but quite often simplified to $\tilde{G}_S = \frac{<|E^{local}_{\omega_p}|^4>}{<|E^{inc}_{\omega_p}|^4>}$. The subscripts $S$ and $p$ stand for Stokes and pump, respectively, so that $|E^{local}_{\omega_s}|$ is the Stokes field amplitude, $|E^{local}_{\omega_p}|$ is the local pump field amplitude normalized by the incident field amplitude, $|E^{inc}_{\omega_p}|$. The brackets represent spatial averages in proximity of the surface. Although $G_S$ is considered more accurate than $\tilde{G}_S$ (for example, $\tilde{G}_S$ requires the frequency of the pump to nearly coincide with the emitted Stokes frequency), both definitions fall short because they: (i) neglect all dynamical aspects of the interaction that relate to the spatial distribution of nonlinear dipoles; (ii) overlook the possibility that the medium may become inverted should local field intensities be sufficiently large; and (iii) ignore phase matching properties and interaction of the generated fields [4]. For a more extended discussion of some of the issues that arise with these definitions of Raman gain we refer the reader to reference [1]. A discussion of additional concerns about $\tilde{G}_S$ may be found in reference [3], where correction terms are derived to improve its accuracy. We simply note here that a full-wave approach that considers the dynamics of the fields can yield significant differences compared to predictions based on the $\tilde{G}_S$ factor alone [1, 5, 6].

SERS studies are usually carried out by either performing a simple calculation of linear field profiles to estimate gain, typically only for one generated field, or in more complicated schemes by assuming that only a small portion of the medium is excited while most particles remain in the ground state. This condition is achieved if the local particle density is much larger than the local photon density at all times, circumstances that are equivalent to assuming the system may be described by classical oscillators [7, 8], i.e. only two components (the complex polarization) of the Bloch vector [8]. Under those circumstances, recent predictions have suggested that in the spontaneous emission regime a structure similar to that depicted in Fig. 1(a) can enhance the local field by two orders of magnitude, followed by improvements of Stokes and anti-Stokes conversion efficiencies by more than seven orders of magnitudes compared to cases without the metal nanostructure [1].

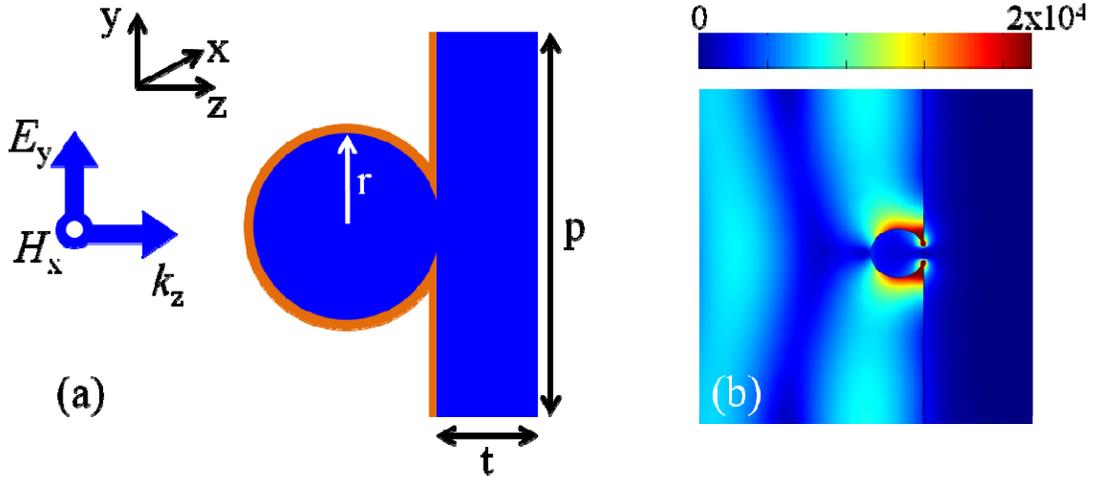

**Fig. 1:** (a) Unit cell of a generic 1D periodic array composed of metal nanowires placed near a metal substrate. (b) An illustration of extreme environment for the geometry of Fig. 1(a): *aluminum nanowire* of radius $r$=20 nm attached to an *aluminum substrate*. Incident wavelength is $\lambda$=220 nm. The local field is amplified by a factor of $2\times10^4$. Incident fields with peak powers in the W/cm$^2$ range can achieve local field intensities of order 1GW/cm$^2$. In this example we use aluminum because silver does not perform as well in the wavelength range of interest.

In this paper we show that by using sub-picosecond pulses and by allowing the system to depart from the ground state, a number of effects never discussed before in this context are triggered: (i) population inversion may occur within the hot spots; (ii) Stokes and anti-Stokes gains saturate; (iii) the transition to what we refer to as the fully nonlinear or stimulated regime occurs in a switch-like fashion reminiscent of a bistable state, as predicted conversion efficiencies jump by an additional *twelve orders of magnitude* beyond the enhancement achievable in the spontaneous emission regime. These results imply that conditions readily achieved in plasmonic nanostructures may require a description of the dynamics of the *full* Bloch vector. For example, in Fig. 1(b) we illustrate the electric field amplitude |**E**| around an *aluminum* nanowire placed near an *aluminum* substrate. Calculations using the dielectric function of aluminum show that the local field is enhanced in excess of four orders of magnitude, so that even ambient light may be capable of inducing huge local fields: a mere fraction of a W/cm$^2$ may strengthen to nearly a GW/cm$^2$, while a focused beam could easily surpass the TW/cm$^2$.

In what follows we highlight the most salient points of the model outlined in reference [1], and then proceed to present new predictions that take into account electronic inversion. In the wavelength range longer than 300 nm the metal sections of the structures, assumed to be composed of silver for the geometry of Fig. 1(a), are described by a combined Drude-Lorentz model as follows:

$$\ddot{\mathbf{P}}_f + \tilde{\gamma}_f \dot{\mathbf{P}}_f = \frac{n_{0,f} e^2}{m_f^*} \left(\frac{\lambda_0}{c}\right)^2 \mathbf{E} + \frac{5}{3}\frac{E_F}{m_f^* c^2} \nabla(\nabla \cdot \mathbf{P}_f), \qquad (1)$$

$$\ddot{\mathbf{P}}_1 + \tilde{\gamma}_{01} \dot{\mathbf{P}}_1 + \tilde{\omega}_{0,1}^2 \mathbf{P}_1 = \frac{n_{0,1} e^2}{m_b^*} \mathbf{E}, \qquad (2)$$

$$\ddot{\mathbf{P}}_2 + \tilde{\gamma}_{02} \dot{\mathbf{P}}_2 + \tilde{\omega}_{0,2}^2 \mathbf{P}_2 = \frac{n_{0,2} e^2}{m_b^*} \mathbf{E}, \qquad (3)$$

where $E_F$ is the Fermi energy of the metal, $\tilde{\omega}_{p,f}, \tilde{\gamma}_f$ are the scaled free electron plasma frequency and damping coefficient, $\tilde{\omega}_{p,1}, \tilde{\omega}_{0,1}, \tilde{\gamma}_{0,1}$ and $\tilde{\omega}_{p,2}, \tilde{\omega}_{0,2}, \tilde{\gamma}_{0,2}$ are the scaled plasma frequencies, resonance frequencies, and damping coefficients that account for inner-core, *d*-orbital (e.g., 4-*d* for Ag and 5-*d* for Au) contributions to the dielectric constant. $n_{0,f}, n_{0,1}, n_{0,2}$ are the free (subscript *f*) and two distinct, bound electron densities (subscripts *1* and *2*); for simplicity we assume that the effective electron masses in the conduction and valence bands of the metal are $m_f^* = m_b^* = m_e$, and that $n_{0,f} \approx n_{0,1} \approx n_{0,2}$. The total linear, metal polarization is then given by $\mathbf{P}_{T,metal} = \mathbf{P}_f + \mathbf{P}_1 + \mathbf{P}_2$. The last term on the right hand side of Eq. (1) imparts nonlocal characteristics to the portion of the dielectric constant arising from free electron gas pressure, i.e. $\varepsilon = \varepsilon(\omega, \mathbf{k})$. In this context the nonlocal term blue-shifts the plasmonic band structure, as shown in Fig. 2(a), inducing dynamic changes to the effective, complex dielectric permittivity.

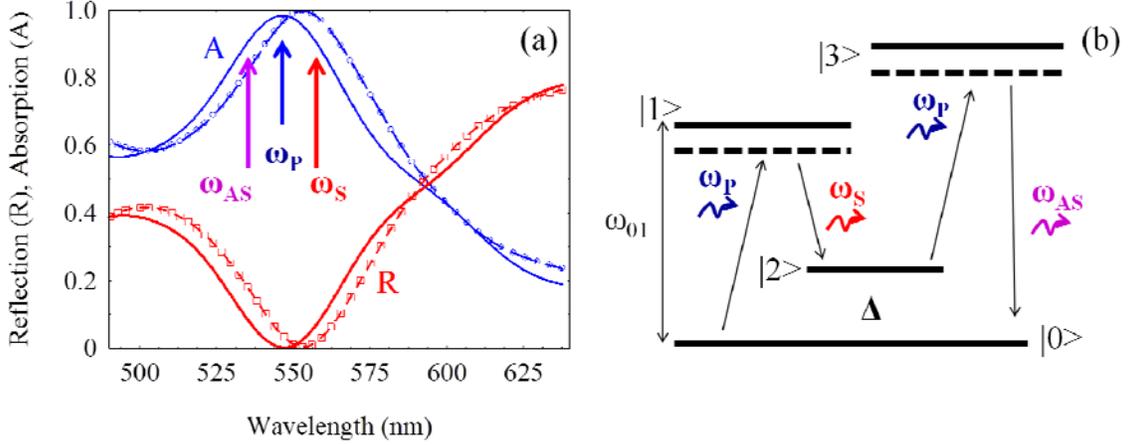

**Fig. 2**: (a) Normalized linear reflection (R) and absorption (A) spectra for the structure in Fig. 1(a) with $r=36$ nm, $t=50$ nm, $p=180$ nm, with (solid) and without (markers) nonlocal effects, obtained by integrating Eqs. (1-3) together with Maxwell's equations, using an incident pulse a few femtoseconds in duration. (b) First Stokes and first anti-Stokes transitions exemplified on a four-level scheme that reduces to an effective two level atom, with coherence between levels |0> and |2>.

The scaled, dimensionless coordinates that appear in Eqs. (1-3) in the form of time derivatives and the operator $\nabla = \frac{\partial}{\partial \tilde{y}} \mathbf{j} + \frac{\partial}{\partial \xi} \mathbf{k}$ are: $\xi = z/\lambda_0$, $\tilde{y} = y/\lambda_0$, and $\tau = ct/\lambda_0$, where the scaling factor is chosen to be $\lambda_0 = 1$ μm. With reference to the coordinate system shown in Fig. 1(a), a TM-polarized field initially propagating in the $z$-direction may be defined as:

$$\mathbf{E} = E_y \mathbf{j} + E_z \mathbf{k} = \left( \sum_l \mathcal{E}_{l,y}(\mathbf{r},t) e^{i(k_{z,l}z - \omega_l t)} + \mathcal{E}_{l,y}^*(\mathbf{r},t) e^{-i(k_{z,l}z - \omega_l t)} \right) \mathbf{j} + \left( \sum_l \mathcal{E}_{l,z}(\mathbf{r},t) e^{i(k_{z,l}z - \omega_l t)} + \mathcal{E}_{l,z}^*(\mathbf{r},t) e^{-i(k_{z,l}z - \omega_l t)} \right) \mathbf{k}$$

$$\mathbf{H} = H_x \mathbf{i} = \left( \sum_l \mathcal{H}_{l,x}(\mathbf{r},t) e^{i(k_{z,l}z - \omega_l t)} + \mathcal{H}_{l,x}^*(\mathbf{r},t) e^{-i(k_{z,l}z - \omega_l t)} \right) \mathbf{i}$$

, (4)

where $\mathcal{E}_{l,y}(\mathbf{r},t)$, $\mathcal{E}_{l,z}(\mathbf{r},t)$ and $\mathcal{H}_{l,x}(\mathbf{r},t)$ are general envelope functions that are allowed to vary rapidly in space and time, $k_{z,l}$ and $\omega_l$ are carrier wave-vectors and frequencies for the $l^{th}$ field, where $l$ enumerates anti-Stokes, pump, and Stokes fields. Beginning with the Heisenberg equations, in the picosecond and sub-picosecond regimes one may neglect super-radiant and sub-radiant decay rates and frequency detunings, and write the Bloch equations of motion for the Raman-active portion of the medium for the $k^{th}$ Cartesian component as follows [1, 9-11]:

$$\dot{W}_k = \frac{\mu_0^2 \lambda_0}{\hbar^2 \omega_{01} c} \left( Q_k^* \sum_{l=1}^{2} \mathcal{E}_{l,k}(\mathbf{r},\tau) \mathcal{E}_{l+1,k}^*(\mathbf{r},\tau) + c.c. \right), \tag{5}$$

$$\dot{Q}_k = -\frac{\mu_0^2 \lambda_0}{2\hbar^2 \omega_{01} c} \left( \sum_{l=1}^{2} \mathcal{E}_{l,k}(\mathbf{r},\tau) \mathcal{E}_{l+1,k}^*(\mathbf{r},\tau) \right) W_k + F_k(\mathbf{r},\tau). \tag{6}$$

The summations are such that $\mathcal{E}_{1,k}(\mathbf{r},\tau)$, $\mathcal{E}_{2,k}(\mathbf{r},\tau)$, $\mathcal{E}_{3,k}(\mathbf{r},\tau)$ are the anti-Stokes, pump, and Stokes fields, respectively, as we will see in the example below. $Q_k \equiv -i\rho_{02}^k$ is the nonlinear polarization, $i \equiv \sqrt{-1}$, $\rho_{jl}^k$ are density matrix elements for the system in Fig. 2(b), $W_k \equiv \rho_{22}^k - \rho_{00}^k$ is the electronic inversion, $\hbar\omega_{01}$ is the energy that separates the ground state from the uppermost, unpopulated state, as shown in Fig. 2(b). $F_k(\mathbf{r},\tau)$ is a Gaussian, white-noise source term with known statistical properties [11] that dominates the spontaneous emission regime. This means that an incident pump field will have little impact on conversion efficiencies until a certain threshold is reached. At that point the field-dependent terms in Eqs. (5) and (6) begin to contribute by triggering a fully nonlinear regime and stimulated processes. The dipole moment on the y-z plane is $\boldsymbol{\mu} = \mu_y \mathbf{j} + \mu_z \mathbf{k}$, where we choose $\mu_y = \mu_z = \mu_0$. The $k^{th}$ Cartesian component of the polarization of the Raman active medium is given by [1, 10]:

$$P_k = \chi_L E_k + i\chi_L \left( Q_k E_k e^{-i\delta t} - Q_k^* E_k e^{i\delta t} \right). \tag{7}$$

Here $E_k = E_k(\mathbf{r},\tau)$ is the $k^{th}$ component of the real field in Eq. (4), and $\chi_L = 4N\mu_0^2/\hbar\omega_{01}$ is the background, linear susceptibility of the medium. Finally, the nonlinear polarization terms that oscillate at the pump, Stokes and anti-Stokes frequencies are [10]:

$$\begin{aligned}
\mathscr{P}_{\omega_p,k}^{NL} &= i\chi_L \left[ Q_k \mathscr{E}_{S,k} - Q_k^* \mathscr{E}_{AS,k} \right] \\
\mathscr{P}_{\omega_S,k}^{NL} &= -i\chi_L Q_k^* \mathscr{E}_{P,k} \\
\mathscr{P}_{\omega_{AS},k}^{NL} &= i\chi_L Q_k \mathscr{E}_{P,k}
\end{aligned} \qquad (8)$$

where $\mathscr{E}_{P,k}$, $\mathscr{E}_{S,k}$, $\mathscr{E}_{AS,k}$ are the generic pump, Stokes, and anti-Stokes field envelopes, respectively, used in Eq. (4), and $\mathscr{P}_{\omega_p,k}^{NL}$, $\mathscr{P}_{\omega_S,k}^{NL}$ and $\mathscr{P}_{\omega_{AS},k}^{NL}$ are the corresponding nonlinear polarizations whose components are added to the total linear polarization that arises from Eqs. (1-3). Although we have limited our expansions to include only one pump and the first Stokes and anti-Stokes fields, more pumps and higher order red- and blue-shifted fields may easily be included to reflect more complex conditions.

It is instructive to take a brief look at the qualitative aspects of the first order solutions of coupled Eqs. (5) and (6) at a fixed oscillator's site. Assuming only a first Stokes and a first anti-Stokes fields are generated, and neglecting the noise source term $F_k(\mathbf{r},\tau)$, Eqs. (5-6) become:

$$\begin{aligned}
\dot{W}_k &= \frac{\mu_0^2 \lambda_0}{\hbar^2 \omega_{01} c} \left[ Q_k^* \left( \mathscr{E}_{A,k} \mathscr{E}_{P,k}^* + \mathscr{E}_{P,k} \mathscr{E}_{S,k}^* \right) + Q_k \left( \mathscr{E}_{A,k}^* \mathscr{E}_{P,k} + \mathscr{E}_{P,k}^* \mathscr{E}_{S,k} \right) \right] \\
\dot{Q}_k &= -\frac{\mu_0^2 \lambda_0}{2\hbar^2 \omega_{01} c} \left( \mathscr{E}_{A,k} \mathscr{E}_{P,k}^* + \mathscr{E}_{P,k} \mathscr{E}_{S,k}^* \right) W_k
\end{aligned} \qquad (9)$$

Using a simple trapezoidal rule, at $t+\delta t$ and position $\mathbf{r}_0$ the approximate solutions of Eqs. (9) are:

$$\begin{aligned}
W_k(\mathbf{r}_0, t+\delta t) &\approx W_k(\mathbf{r}_0, t) + \frac{\mu_0^2 \lambda_0}{\hbar^2 \omega_{01} c} \left( Q_k^*(\mathbf{r}_0,t) F(\mathbf{r}_0,t) + Q_k(\mathbf{r}_0,t) F_k^*(\mathbf{r}_0,t) \right) \delta t \\
Q_k(\mathbf{r}_0, t+\delta t) &\approx Q_k(\mathbf{r}_0, t) - \frac{\mu_0^2 \lambda_0}{2\hbar^2 \omega_{01} c} F_k(\mathbf{r}_0,t) W_k(\mathbf{r}_0,t) \delta t
\end{aligned} \qquad (10)$$

where $F_k(\mathbf{r}_0,t) = \mathscr{E}_{A,k}(\mathbf{r}_0,t) \mathscr{E}_{P,k}^*(\mathbf{r}_0,t) + \mathscr{E}_{P,k}(\mathbf{r}_0,t) \mathscr{E}_{S,k}^*(\mathbf{r}_0,t)$. It is evident that the polarization $Q_k(\mathbf{r}_0, t+\delta t)$ that folds back into Eqs. (8) cannot be characterized as a simple third order, intensity-dependent process. With this in mind, we consider a silver nanowire/film system with $r$=36 nm, $t$=50 nm, and $p$=180 nm; both the nanowire and the substrate are coated with a ~5 nm-thick layer of Raman-active material, as shown in Fig. 1(a). Incident pulses of varying duration and peak intensity have a carrier wavelength of approximately $\lambda_p = 550$ nm, where linear reflection (absorption) is minimized (maximized), as shown on Fig. 2(a). Stokes and anti-Stokes fields are tuned to $\lambda_S = 565$ nm and $\lambda_{AS} = 535$ nm, respectively. Typical dipole moments of diatomic molecules range from 1 to 10 Debyes (symbol D; $1D = 10^{-10}$ esu $\times 10^{-8}$ cm), to ~$4.5 \times 10^4$ D for DAST crystals 350 nm in size [12]. We choose a dipole moment $\mu_0 = 30 e a_0$, where $e$ is the electron charge and $a_0$ is the Bohr radius, so that $\mu_0 \sim 72$ D. The $\mu_0^2$ dependence of the coupling constants in Eqs. (5-6) helps keep peak, input intensities reasonably large, and computation times manageably reasonable. Finally, we choose a molecular density consistent with a gaseous substance having a background index $n_L \approx 1.001$. For additional details on geometrical considerations and method of integration we refer the reader to reference [1].

In Fig. 3(a) we depict reflected Stokes conversion efficiency as a function of incident peak intensity for pulse durations that range from 125 fs to 1 ps (full width at half maximum of the Gaussian intensity profile). Incident pulses with peak power below ~1 GW/cm$^2$ yield conversion efficiencies that are enhanced by roughly seven orders of magnitude compared to conversion efficiencies calculated without the metal nanostructure, i.e. a free-standing, 5 nm-thick Raman layer (horizontal, dashed black curve). Although it is not surprising to find that pulses of longer duration have lower nonlinear thresholds, the spectral width of a 125 fs pulse (~10 nm bandwidth) is already well-contained within the resonance of Fig. 2(b). Therefore, conversion efficiency and the degree of inversion may be somewhat sensitive to: (i) pulse area, intended in the sense of: Area $\propto \int_{-\infty}^{\infty} E(\tau) d\tau$ [8], although it is not known how π-pulses impact the effective two-level system we are considering in an environment with strong feedback; (ii) exact tuning of Stokes and anti-Stokes signals with respect to the pump and the plasmonic band structure; and perhaps more importantly, (iii) the simultaneous presence of multiple generated fields, as we will see below. We thus find that the transition region between spontaneous and stimulated regimes sharpens for longer pulses and displays a remarkable jump of more than twelve orders of magnitude in conversion efficiency relative to the spontaneous emission regime, and nearly twenty orders of magnitude compared to the 5 nm layer alone. Anti-Stokes efficiencies show similar behavior.

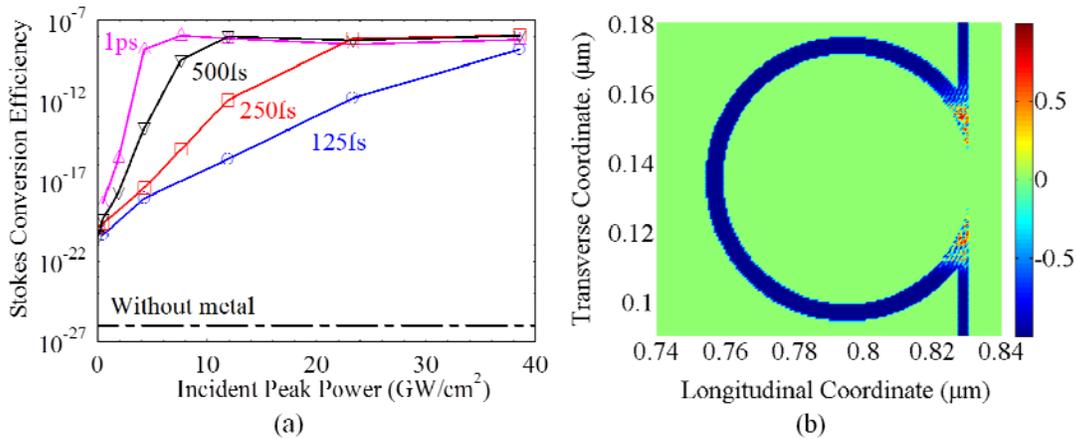

**Fig. 3**: (a) Reflected Stokes efficiency vs. incident peak power for pulse durations as shown. (b) Spatial distribution of the inversion associated with the longitudinal component of the field. The dark blue regions represent a Raman medium in the ground state. The medium is inverted in the yellow and red spots, where the local field is largest.

At the same time, the hot spots become inverted ($W \geq 0$) in the regions above and below the nanowire-substrate contact point – Fig. 3(b). Unlike a classical harmonic oscillator, Fig. 3(b) shows that the excited state |2> in Fig. 2(b) can in fact store a significant amount of energy, with the additional consequence that fewer pump photons partake in the conversion process. The instability, or switching, may arise from either local field effects or the strong feedback present in the system. It is well-known that, when inserted in a cavity, a two-level atom displays optical bistability [13]. As for local field effects in two-level systems, the recognition that the local field $E_L$ and the Maxwell field $E$ are related via the polarization $P$ ($E_L = E + 4\pi P/3$), leads to a dynamic frequency detuning proportional to the inversion [14], that in turn can invert the system in switch-like fashion [15] without the need for a cavity, a phenomenon known as intrinsic optical bistability. Both effects are simultaneously present.

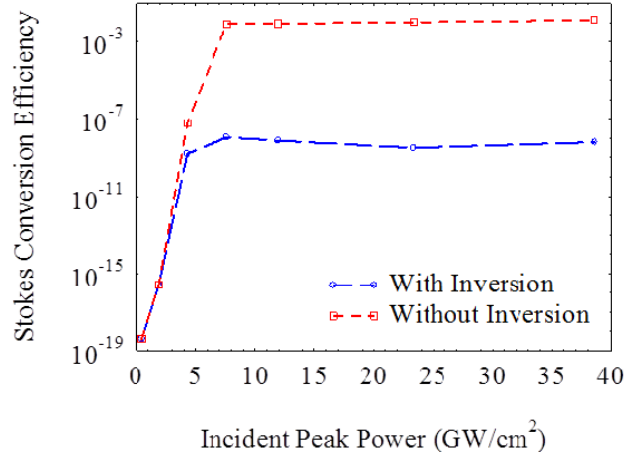

**Fig. 4**: Reflected Stokes efficiencies vs. incident peak power for 1ps pulses, with (long dashes, empty circles) and without (short dashes, empty squares) inversion. Inversion inhibits efficiency as the medium stores energy and saturation occurs.

In Fig. 4 we plot conversion efficiency for 1ps pulses, with and without the contribution of inversion. Although both curves display switching dynamics, the figure suggests that in the absence of inversion Raman gain saturates as the pump becomes depleted. This process may be accompanied by a degree of self- and cross-phase modulation and spectral broadening of the fields that does not occur if inversion were allowed to take part in the process. If in fact the inversion is allowed to evolve in time, the conversion process is inhibited as progressively fewer atoms are available for the production of either Stokes or anti-Stokes photons. As the peak intensity is increased further the extra energy supplied simply inverts additional portions of the medium in the neighborhood of the metal contact point.

Finally, we look at the consequences of the simultaneous presence of multiple generated fields. The summations in Eqs. (5), (6), and (9) suggest that if red- and blue-shifted fields are present concurrently, including higher order Stokes and anti-Stokes fields, they can interfere and change the evolution of the atomic variables to a degree that depends on the magnitude of the dipole moment for each process. Here we have assumed that the coupling coefficients (i.e. the size of the dipole moment) are roughly similar for all processes. In turn, this interference affects the evolution of the fields according to the specific nature of each source term, i.e. Eqs. (8). This point is illustrated in Fig. (5), where we compare Stokes conversion efficiency with and without the anti-Stokes field, for incident 500 fs pulses. In general, suppressing the anti-Stokes channel makes all pump photons available exclusively to the Stokes process − see level diagram in Fig. 2(b). Consistent with this, in the nonlinear regime Stokes emissions evolve more rapidly in time, requiring shorter pulses and saturating for input intensities that are lower by at least one order of magnitude compared to the case that includes both Stokes and anti-Stokes fields. The nonlinear threshold is then lowered to a few tens of MW/cm$^2$. Similar arguments hold for either multiple red-shifted or multiple blue-shifted fields.

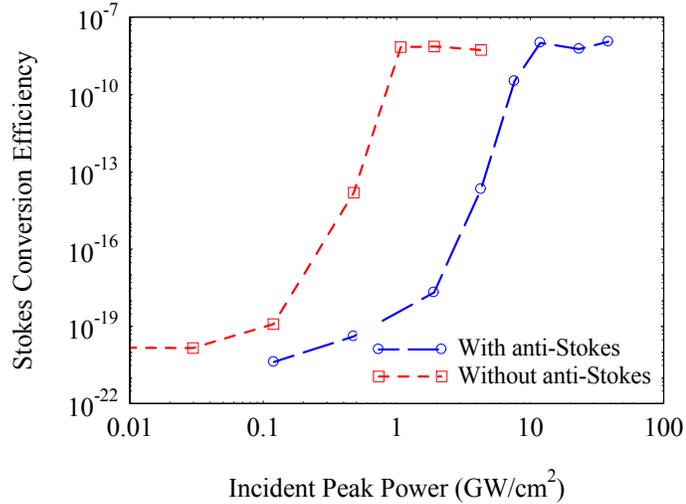

**Fig. 5**: Reflected Stokes efficiency vs. incident peak power for 500fs pulses, with (long dashes, empty circles) and without (short dashes, empty squares) anti-Stokes channel. The anti-Stokes field interferes with Stokes sources, modifies the evolution of the atomic variable, and changes the dynamics of the fields themselves. If the anti-Stokes field is absent more photons are available to drive the Stokes process, lowering nonlinear thresholds by approximately one order of magnitude.

As we have seen, the structure we are considering enhances the local field by two orders of magnitude, a modest amount compared, for example, to the structure of Fig. 1(b). Combining this with our chosen dipole moment yields a nonlinear threshold of approximately 1 GW/cm$^2$ − Fig. 3(a) and Fig. 4. Then, within each hot spot the local field intensity can reach values of several tens of TW/cm$^2$, which may suffice to trigger ionization effects and plasma formation that we have not considered here, but that we realize could introduce additional, unforeseen effects. Conversely, a mere doubling of the effective dipole moment pushes the nonlinear threshold for Stokes and anti-Stokes generation down into the hundreds of MW/cm$^2$, while a dipole moment 10 times larger lowers the nonlinear threshold for this particular structure down to a few MW/cm$^2$. Furthermore, a number of appropriate geometrical and/or material modifications could easily improve local field enhancement by a few orders of magnitude (e.g. Fig. 2(b)), thus lowering nonlinear thresholds further, avoiding field ionization effects and placing the nonlinear regime within reach of many experimental set-ups. This notwithstanding, the nonlinear threshold decreases by at least one order of magnitude if only one field is generated (Fig. 5).

To summarize, we have shown that field localization effects within the hot spots of a Raman medium generally require full accounting of the dynamics of the entire Bloch vector. As the medium becomes steadily inverted, the local photon density can approach and even surpass particle density, causing the Stokes and anti-Stokes conversion channels to clamp down and the efficiency to saturate. Nevertheless, our system, in no way optimized, displays a remarkably large switch-like transition between the spontaneous and stimulated regimes that spans nearly twelve orders of magnitude.

**Acknowledgement**

M. Grande thanks the U.S. Army International Technology Center Atlantic for financial support (W911NF-12-1-0292). This research was performed while the authors M. A. Vincenti, D. de Ceglia and J. W. Haus held National Research Council Research Associateship awards at the U. S. Army Aviation and Missile Research Development and Engineering Center.